\documentclass[11pt,amsmath,amssymb,nofootinbib]{revtex4}
\usepackage{amsmath}
\usepackage{dcolumn}
\usepackage{bm}
\usepackage[all, knot]{xy}
\xyoption{arc}
\begin{document}
\markboth{Amelino-Camelia, Gubitosi, Marcian\`o, Martinetti,
Mercati, Pranzetti, Tacchi }{First results of the Noether theorem
for Hopf-algebra spacetime symmetries}

\title{First results of the Noether theorem
 for Hopf-algebra spacetime
symmetries\footnote{Based in part on the lecture given by G.A.-C.~
at the 21st Nishinomiya-Yukawa Memorial Symposium {\it
Noncommutative geometry and quantum spacetime in physics}, but
updated on the basis of the related results more recently obtained
in Refs.~[2,3,4]}}

\author{Giovanni \textsc{Amelino-Camelia}}
\affiliation{Dipartimento di Fisica \\
Universit\`a degi Studi di Roma ``La Sapienza"\\
and Sez.~Roma1 INFN\\
Ple Moro 2, Roma 00185, Italy}
\author{Giulia \textsc{Gubitosi}}
\affiliation{Dipartimento di Fisica \\
Universit\`a degi Studi di Roma ``La Sapienza"\\
and Sez.~Roma1 INFN\\
Ple Moro 2, Roma 00185, Italy}
\author{Antonino \textsc{Marcian\`o}}
\affiliation{Dipartimento di Fisica \\
Universit\`a degi Studi di Roma ``La Sapienza"\\
and Sez.~Roma1 INFN\\
Ple Moro 2, Roma 00185, Italy}
\author{ Pierre \textsc{Martinetti}}
\affiliation{Dipartimento di Fisica \\
Universit\`a degi Studi di Roma ``La Sapienza"\\
and Sez.~Roma1 INFN\\
Ple Moro 2, Roma 00185, Italy}
 \author{Flavio \textsc{Mercati}}
 \affiliation{Dipartimento di Fisica \\
Universit\`a degi Studi di Roma ``La Sapienza"\\
and Sez.~Roma1 INFN\\
Ple Moro 2, Roma 00185, Italy}
  \author{Daniele
\textsc{Pranzetti}} \affiliation{Dipartimento di Fisica\\
Universit\`a di Roma Tre\\
Via Vasca Navale 84, 00146 Roma, Italy}
 \author{Ruggero Altair \textsc{Tacchi}}
\affiliation{Physics Department \\
University of California,\\
Davis, CA 95616, US}

\begin{abstract}
\begin{center}
{\bf Abstract}
\end{center}
We summarize here the first results obtained using a technique we
recently developed for the Noether analysis of Hopf-algebra
spacetime symmetries, including the derivation of conserved charges
for field theories in noncommutative spacetimes of canonical or
$\kappa$-Minkowski type.
\end{abstract}
\maketitle
\newpage\baselineskip12pt plus .5pt minus .5pt \pagenumbering{arabic} %
\pagestyle{plain}

\section{Introduction}
In these notes we summarize the first results obtained using a
technique for the Noether analysis of Hopf-algebra spacetime
symmetries which we developed in
Refs.~\cite{k-Noether,NopureBoost,k-Noether5D,theta-Noether}. The
mathematics of Hopf algebras is considered as a promising candidate
in the search of a formalism compatible with the idea of
Planck-scale deformation of spacetime symmetries, possibly in the
sense of ``doubly-special relativity"~\cite{gacdsr,kowadsr,leedsr}.
However, progress in this direction has been for a long time stalled
by our inability to establish what is the fate of
physical/observable aspects of spacetime symetries in the
Hopf-algebra framework. It is legitimate to hope that the Noether
charges obtained through our Noether analyses will prove valuable
for the debate on these issues.

The Hopf-algebra spacetime symmetries we
analyzed~\cite{k-Noether,NopureBoost,k-Noether5D,theta-Noether} are
relevant for field theories constructed in canonical noncommutative
spacetimes, with the characteristic noncommutativity of coordinates
given by
\begin{equation}
[\hat{x}^\mu,\hat{x}^\nu]=i \lambda^2 \Theta^{\mu\nu} \equiv i
\theta^{\mu\nu} ~,\label{thetacommutationrelation}
\end{equation}
or in the so-called $\kappa$-Minkowski noncommutative
spacetime~\cite{majrue,kpoinap}, a Lie-algebra~\cite{wessDefinizio}
noncommutative spacetime with
\begin{eqnarray}
[\hat{x}_j,\hat{x}_0]= i \lambda \hat{x}_j
~,~~~~[\hat{x}_k,\hat{x}_j]=0 ~,\label{kmnoncomm}
\end{eqnarray}
where $\hat{x}_0$ is the time coordinate, $\hat{x}_j$ are space
coordinates ($j,k \in \{1,2,3\}$, $\mu,\nu \in \{0,1,2,3\}$),
$\lambda$ is an observer-indepedent~\cite{gacdsr} length scale,
usually expected to be of the order of the Planck length, and
$\Theta^{\mu\nu}$ ($\equiv \theta^{\mu\nu}/\lambda^2$) is a
dimensionless coordinate-independent and
observer-independent\footnote{The earliest
studies~\cite{doplich1994} of noncommutativity with a
coordinate-independent $\theta^{\mu\nu}$ actually adopted an even
richer formalism, in particular attributing to $\theta^{\mu\nu}$
some nontrivial algebraic properties~\cite{doplichRecent}. A large
literature has been devoted to the simplest picture with a
coordinate-independent $\theta^{\mu\nu}$, in which $\theta^{\mu\nu}$
is a (dimensionful) number-valued (observer-dependent)
tensor~\cite{szabo,douglas,susskind}, giving rise to a rather
familiar type of break down of spacetime symmetries (emergence of a
preferred frame). The possibility we considered, the one of a
$\theta^{\mu\nu}$ that is a number-valued
{\underline{observer-independent}} matrix, was developed more
recently, mostly through the works reported in
Refs.~\cite{chaichaTwist,wessTwist,balaTwist}.} matrix.

In pursuing the objective of deriving conserved charges from the
relevant Hopf-algebra spacetime symmetries we
stumbled~\cite{k-Noether,NopureBoost,k-Noether5D,theta-Noether} upon
the striking (though, {\it a posteriori}, obvious) realization that
the symmetry-transformation parameters should not commute with the
spacetime coordinates. And the form of the relevant commutation
relations is such that certain types of ``pure transformations" are
not allowed. Canonical noncommutative spacetimes admit pure
translation transformations, but any transformation involving a
Lorentz-sector component must also have a nonvanishing translation
component. Similarly in the $\kappa$-Minkowski case pure
translations and pure space rotations are allowed, but any
transformation involving a boost component must also have a
nonvanishing space-rotation component.

\section{$\kappa$-Minkowski noncommutative spacetime}
Let us start with the analysis of $\kappa$-Minkowski spacetime,
following the results we reported in
Refs.~\cite{k-Noether,NopureBoost}. For simplicity we shall focus
here on the ordering convention\footnote{As discussed in
Refs.~\cite{k-Noether,NopureBoost,aadluna} one can legitimately
adopt other ordering conventions, but for brevity we shall here
neglect this possibility.} such that a generic function of the
noncommuting $\kappa$-Minkowski coordinates is written as a Fourier
sum of ``time-to-the-right-ordered" exponentials~\cite{aadluna}:
\begin{equation}
\Phi (\hat{x}) = \int d^4 k ~ \tilde{\Phi}(k) ~ e^{i \vec{k} \cdot
\vec{\hat{x}}} e^{- i k_0 \hat{x}_0}~, \label{FourTrans}
\end{equation}
where the Fourier parameters $k_\mu$ are commutative and $\int d^4
k$ is an ordinary integral.

We shall consider a (classical) field theory in $\kappa$-Minkowski
which is invariant under transformations generated by ``classical
action" translation and space-rotation generators:
\begin{equation}
P_\mu \left(  e^{i \vec{k} \cdot \vec{\hat{x}}} e^{- i k_0
\hat{x}_0} \right) = k_\mu ~ e^{i \vec{k} \cdot \vec{\hat{x}}} e^{-
i k_0 \hat{x}_0}, ~ R_j \left(  e^{i \vec{k} \cdot \vec{\hat{x}}}
e^{- i k_0 \hat{x}_0}    \right) = i \varepsilon_{jkl} ~ \hat{x}_k ~
k_l ~ e^{i \vec{k} \cdot \vec{\hat{x}}} e^{- i k_0 \hat{x}_0} ~,
 \label{ActionOnExpAAA}
\end{equation}
and boost generators with the rule of action
\begin{eqnarray}
N_j \left(  e^{i \vec{k} \cdot \vec{\hat{x}}} e^{- i k_0 \hat{x}_0}
\right) = \left[ \hat{x}_j \left(\frac{1 - e^{- 2 \lambda k_0}}{2
\lambda} + \frac{\lambda}{2} |\vec{k}|^2 \right) - \hat{x}_0 k_j
\right] e^{i \vec{k} \cdot \vec{\hat{x}}} e^{- i k_0 \hat{x}_0}   ~,
\label{ActionOnExpBBB}
\end{eqnarray}
which can be obtained~\cite{aadluna} by imposing that the $N_j$,
together with the $P_\mu$ and the $R_j$, are compatible with the
requirements for a Hopf algebra. This turns out to be the
$\kappa$-Poincar\'e Hopf algebra (written in the ``bicrossproduct
basis''~\cite{majrue}).

 We shall first argue that the transformations generated by these generators
 require noncommutative transformation parameters, and then perform
 a Noether analysis for the corresponding description of symmetry transformations.

 In some points of the analysis we shall of course resort to 4D and 3D integration
 over the $\kappa$-Minkowski coordinates, which we shall perform consistently
 with~\cite{k-Noether,NopureBoost,aadluna}
\begin{equation}
\int d^4 \hat{x} ~ e^{i \vec{k} \cdot \vec{\hat{x}}} e^{- i k_0
\hat{x}_0} = \delta^{(4)} (k) ~,
\end{equation}
and
\begin{equation}
\int d^3\hat{x} ~ e^{i \vec{k} \cdot \vec{\hat{x}}} e^{- i k_0
\hat{x}_0} = \delta^{(3)} (\vec{k}) e^{- i k_0 \hat{x}_0}~.
\end{equation}

\subsection{Noncommutative transformation parameters} \label{DifferentialCalculus}
In Refs.~\cite{k-Noether,NopureBoost} we sought a description of our
transformations on functions $f(\hat{x})$ of the $\kappa$-Minkowski
coordinates of the familiar type $f \rightarrow f+df$, with
\begin{equation}
df(\hat{x})=i\gamma^\mu P_\mu f(\hat{x}) + i \tau_j N_j f(\hat{x}) +
i \sigma_k R_k f(\hat{x}), \label{differenzialeTot}
\end{equation}
and for the transformation parameters $\gamma^\mu$, $\tau_j$,
$\sigma_k$ we insisted that they should act on spacetime coordinates
by associative (but not necessarily commutative) multiplication. We
further insisted that $d$ satisfies Leibniz rule,
\begin{equation}
d(f(\hat{x})g(\hat{x}))=(df(\hat{x}))g(\hat{x})+f(\hat{x})(dg(\hat{x}))
~, \label{leibnizKappa}
\end{equation}
and this turns out to be a rather nontrivial requirement, as a
result of the fact that from the definitions
(\ref{ActionOnExpAAA}),(\ref{ActionOnExpBBB}) it follows
that\cite{k-Noether,NopureBoost}
\begin{subequations}
\begin{equation}
P_\mu \left[ f(\hat{x}) g(\hat{x}) \right] = \left[ P_\mu f(\hat{x})
\right] g(\hat{x}) + e^{-\lambda P_0 (1-\delta_{\mu_0})} f(\hat{x})
\left[ P_\mu g(\hat{x}) \right]~,
\end{equation}
\begin{equation}
N_j  \left[ f(\hat{x}) g(\hat{x}) \right] = \left[ N_j f(\hat{x})
\right] g(\hat{x}) +  e^{-\lambda P_0} f(\hat{x}) \left[N_j
g(\hat{x}) \right] + \lambda \varepsilon_{jkl} \left[P_k f(\hat{x})
\right] \left[ R_l g(\hat{x}) \right]~,
\end{equation}
\begin{equation}
R_k  \left[ f(\hat{x}) g(\hat{x}) \right] = \left[ R_k f(\hat{x})
\right] g(\hat{x}) + f(\hat{x}) \left[ R_k g(\hat{x}) \right]~,
\end{equation}
\end{subequations}
which reflect the structure of the so-called ``coproduct" rules of
the bicrossproduct basis of the $\kappa$-Poincar\'e Hopf algebra.

Unlike the corresponding transformation parameters for classical
Minkowski spacetime, the $\gamma^\mu$, $\tau_j$ and $\sigma_k$ must
have noncommutative product rules with the coordinates. The
commutation relations of the parameters with the coordinates turn
out (as shown\footnote{Also see Ref.~\cite{kowafreidIIv1}, which
however did not report the correct form of the commutators between
transformation parameters, and was eventually
revised~\cite{kowafreidIIv2}.} in
Refs.~\cite{k-Noether,NopureBoost}) to take the following form:
 \begin{equation}
 \left\{
 \begin{array}{l}
\left[\gamma_0,\hat{x}_\mu \right] = 0 \\
\left[\gamma_j,\hat{x}_\mu \right] = i \lambda \gamma_j \delta_{\mu}^0\\
   \end{array} \right. ~,
\qquad
 \left\{
 \begin{array}{l}
\left[\tau_j,\hat{x}_k \right] = 0 \\
\left[\tau_j,\hat{x}_0 \right] = i \lambda \tau_j
   \end{array} \right.~,
\qquad
 \left\{
 \begin{array}{l}
 \left[ \sigma_j, \hat{x}_k \right] = i \lambda \varepsilon_{jmk} \tau_m \\
 \left[ \sigma_j, \hat{x}_0 \right] = 0
 \end{array} \right. .
  \label{CommParam}
 \end{equation}
 Interestingly these commutators provide an
 (otherwise unexpected) obstruction for the realization of a pure boost.
 In fact, according to the
commutation relations (\ref{CommParam}), for $\tau_m \neq 0 $ (at
least for some $m$) one necessarily has $\left[ \sigma_j, \hat{x}_k
\right] \neq 0 $ (at least for some $j,k$ combination), so that at
least for some $j$ one must have $ \sigma_j \neq 0$: whenever a
symmetry transformation has a boost component it must also have a
space-rotation component. Clearly no similar obstruction applies to
the cases of a pure translation or a pure space rotation: whenever
$\tau_m =0$ one gets $\left[ \sigma_j, \hat{x}_\mu \right] = 0 =
\left[\gamma_\nu,\hat{x}_\mu \right]$, so in turn one may also set
$\sigma_j = 0$ and/or $\gamma_\nu =0$.

\subsection{Noether analysis}
The description of symmetry transformations discussed in the
previous subsection, encouragingly turns out to allow the derivation
of some associated conserved charges. This was verified explicitly
in Refs.~\cite{k-Noether,NopureBoost} for the illustrative example
of a theory of massless scalar fields $\Phi(\hat{x})$ governed by
the Klein-Gordon-like equation of motion
\begin{equation}
 \square_\lambda \Phi (\hat{x}) \equiv \tilde{P}_\mu \tilde{P}^\mu \Phi
  \equiv \left[ -\left( \frac{2}{\lambda} \right)^2 \sinh^2\left( \frac{\lambda P_0}{2}\right)
  +e^{\lambda P_0}|\vec P|^2 \right] \Phi(\hat{x}) =0 ~,\label{equescionovdemoscion}
\end{equation}
which is the most studied~\cite{kpoinap,kowaorder,aadluna}
 theory formulated in  $\kappa$-Minkowski.
The operator $\square_\lambda$ is the ``mass Casimir" of the
$\kappa$-Poincar\'e Hopf algebra, and we adopted the convenient
notation $ \tilde{P}_{0} \equiv
\left(\frac{2}{\lambda}\right)\sinh(\lambda {P}_{0}/2 )$ , $
\tilde{P}_{j} \equiv e^{\lambda {P}_{0}/2} {P}_{j} $.

We of course consider our transformation rules, which for a scalar
field take the form
\begin{eqnarray}
&\hat{x}'_\mu = \hat{x}_\mu + d \hat{x}_\mu =  \hat{x}_\mu+i
(\gamma^\nu P_\nu  +  \tau_j N_j + \sigma_k R_k ) \hat{x}_\mu ~,&
\\
&\Phi'(\hat{x}') -\Phi(\hat{x}) = \Phi'(\hat{x}')- \Phi(\hat{x}') +
\Phi(\hat{x}') - \Phi(\hat{x})   \simeq  \delta \Phi(\hat{x}) + d
\Phi(\hat{x})  =0 ~,&
\end{eqnarray}
with
\begin{equation}
\delta\Phi= -d \Phi \equiv - \left[i \gamma^\mu P_\mu + i\sigma_j
R_j + i\tau_k N_k \right] \Phi ~, \label{variescion}
\end{equation}

It is easy to verify~\cite{theta-Noether} that the equation of
motion (\ref{equescionovdemoscion}) is invariant\footnote{Note that
the mass Casimir ${\square_\lambda}$ commutes with all the
generators $P_\mu$,$R_j$,$N_j$ of the Hopf algebra. We also make the
natural assumption that the transformation parameters $\gamma_\mu$,
$\sigma_j$, $\tau_j$ also commute with ${\square_\lambda}$. In the
later section on covariance of the transformation parameters we will
present an argument showing that this reasonable assumption does
lead to an overall appealing picture.} ($\delta ( {\square_\lambda}
\Phi) = {\square_\lambda} \delta \Phi = 0$) under these
transformations.

Our Noether analysis takes as starting point the action
\begin{equation}
S = \frac{1}{2} \int d^4\hat{x} \, \Phi(\hat{x}) \,
{\square_\lambda}\, \Phi (\hat{x}), \label{azione} \
\end{equation}
from which the equation of motion (\ref{equescionovdemoscion}) can
be obtained variationally~\cite{k-Noether}.

The result of a variation of the action (\ref{azione}) under our
transformation is:
\begin{equation}
\delta S = \frac{1}{2} \int d^4 \hat{x} \,  \tilde{P}^\mu
\left\lbrace  \tilde{P}_\mu \left[ \left( e^{\lambda P_0} \Phi
\right) \delta \Phi \right] - 2  \left( e^{\lambda P_0}
\tilde{P}_\mu \Phi \right) e^{\frac{\lambda}{2} P_0} \delta
\Phi\right \rbrace,\label{variescionovdeacscion}
\end{equation}
where we already restricted the analysis to fields that are
solutions of the equation of motion (which are the ones whose
charges we are interested in), and we used the following property of
the operators  $\tilde P_\mu$:
\begin{equation}
\tilde{P}_\mu \left[ f(\hat{x}) g(\hat{x}) \right] = \left[
\tilde{P}_\mu f(\hat{x})\right] \left[  e^{\frac{\lambda}{2} P_0}
g(\hat{x})\right] + \left[ e^{-\frac{\lambda}{2} P_0}
f(\hat{x})\right] \left[  \tilde{P}_\mu g(\hat{x})\right]  ~.
\end{equation}
Using the observation~\cite{k-Noether}
\begin{equation}
\int d^4 \hat{x} \, e^{\xi P_0} \left[ f(\hat{x}) \right] = \int
d^4\hat{x} \,f(\hat{x}) ~~~~~~~~ \forall \xi  ~,
\end{equation}
and the fact that from the  rules of commutation (\ref{CommParam})
between transformation parameters and spacetime coordinates it
follows that, for a generic function of the coordinates
$f(\hat{x})$, one has $f(\hat{x}) \gamma_\mu = \gamma_\mu
(e^{-\lambda (1-\delta_\mu^0) P_0}f(\hat{x}))$,  $f(\hat{x}) \tau_j
= \tau_j (e^{-\lambda P_0}f(\hat{x}))$ and $[f(\hat{x}) , \sigma_j]
= \lambda\varepsilon_{jlk} \tau_l (P_k f(\hat{x}))$, one can rewrite
$\delta S$ in the following form:
\begin{equation}
\delta S =  \int d^4 \hat{x} \left(i \gamma_\nu P_\mu T^{\mu \nu} +
i \tau_k P_\mu J^\mu_k + i \sigma_j P_\mu K^\mu_j \right),
\label{Quadricorrenti}
\end{equation}
where:
\begin{subequations}
\begin{equation}
T^{\mu \nu}(\hat{x}) =  \frac{1}{2} \left( e^{\lambda (1 -
\delta_\mu^0) P_0} \tilde{P}^\mu \Phi e^{\frac{\lambda}{2} P_0}
P^\nu \Phi - e^{\lambda  (1/2 - \delta_\mu^0) P_0}\Phi \;
\tilde{P}^\mu P^\nu \Phi   \right) ~,
\end{equation}
\begin{eqnarray}
J^\mu_j (\hat{x}) &=& \frac{1}{2} \left(\tilde{P}^\mu \Phi
e^{\frac{\lambda}{2} P_0}  N_j \Phi - e^{-\frac{\lambda}{2} P_0}
\Phi  \tilde{P}^\mu N_j \Phi \right)
+ \nonumber \\
&&+ \frac{\lambda}{2} \varepsilon_{jkl} \left(e^{\lambda P_0}
\tilde{P}^\mu P_k \Phi e^{\frac{\lambda}{2} P_0}  R_l \Phi  -
e^{\frac{\lambda}{2} P_0} P_k \Phi  \tilde{P}^\mu R_l \Phi
\right)~,
\end{eqnarray}
\begin{equation}
K^\mu_j (\hat{x}) =  \frac{1}{2} \left( e^{\lambda P_0}
\tilde{P}^\mu \Phi e^{\frac{\lambda}{2} P_0}  R_j \Phi -
e^{\frac{\lambda}{2} P_0}\Phi \;  \tilde{P}^\mu R_j \Phi   \right).
\end{equation}
\end{subequations}
And one can explicitly verify~\cite{k-Noether,NopureBoost} that the
ten charges obtained by spatial integration of the $T^{0
\mu}(\hat{x})$, $J^0_j(\hat{x})$ and the $K^0_j(\hat{x})$,
\begin{eqnarray}
Q^{P}_\mu \equiv \int d^3 \hat{x} \, T^0_\mu(\hat{x}) ~, \qquad
Q^{N}_j \equiv \int d^3 \hat{x} \, J^0_j(\hat{x}) ~, \qquad Q^{R}_j
\equiv \int d^3 \hat{x} \, K^0_j(\hat{x}) ~,
\end{eqnarray}
are time independent and can be conveniently written, in terms of
the Fourier transform $\tilde{\Phi}(k)$ of a field $\Phi({\hat{x}})$
solution of the equation of motion, as follows:\footnote{Here and in
the following we will sometimes use the convenient compact notations
$\tilde{k}_0 =2/\lambda \sinh \left( \lambda  k_0  / 2 \right)$,
 $\tilde{k}_j = e^{\frac{\lambda}{2} k_0} k_j$, $\tilde{k}^2 = \tilde{k}_\mu \tilde{k}^\mu$ .}:
\begin{subequations}
\begin{equation}
Q^{P}_\mu = \frac{1}{2} \int d^4 k  \, \frac{\tilde{k}_0}{\left|
\tilde{k}_0\right|} \, e^{ (2 - \delta_\mu^0) \lambda k_0} \,
\tilde{\Phi}(-k_0 ,-e^{\lambda k_0} \vec{k}) k_\mu   \tilde{\Phi}(k)
\delta(\tilde{k}^2) ~,
\end{equation}
\begin{eqnarray}
Q^{N}_j &=& \frac{i}{2} \int   d^4 k \,
\frac{\tilde{k}_0}{|\tilde{k}_0|} e^{ \lambda k_0}
 \tilde{\Phi}(-k_0 ,-e^{\lambda k_0} \vec{k})  \delta(\tilde{k}^2)  \left\{   k_j  \frac{\partial  \tilde{\Phi}(k)  }{\partial k_0} +\right. \nonumber \\
&& \qquad \qquad \left. +  \frac{\partial}{\partial k_j}\left[
\left( \frac{1-e^{-2\lambda k_0}}{2 \lambda} - \frac{\lambda}{2}
|\vec{k}|^2  \right) \tilde{\Phi}(k)\right] - \lambda k_j
\tilde{\Phi}(k)  \right\} ~,
\end{eqnarray}
\begin{equation}
Q^{R}_j = \frac{i}{2} \int d^4 k  \, \frac{\tilde{k}_0}{\left|
\tilde{k}_0\right|} \, e^{ 2 \lambda k_0} \, \tilde{\Phi}(-k_0
,-e^{\lambda k_0} \vec{k}) \varepsilon_{jlm} k_m \frac{\partial
\tilde{\Phi}(k)}{\partial k_l}  \delta(\tilde{k}^2) ~.
\label{AllCharges}
\end{equation}
\end{subequations}

\section{Translation transformations and a 5D differential calculus}
Intriguingly, the Noether analysis reported in the previous section
is somehow related to the structure of a known 4D differential
calculus for 4D $\kappa$-Minkowski: it is easy to
verify~\cite{k-Noether} that the translation-transformation
parameters have commutators with the $\kappa$-Minkowski coordinates
(which we gave in (\ref{CommParam})) that exactly reproduce the
commutators between elements of the relevant 4D calculus and
coordinates.
 In Ref.~\cite{k-Noether5D} (also see Ref.~\cite{kowafreidIIv2})
 we followed the same procedure of analysis for
 a description of translation
transformations in 4D $\kappa$-Minkowski analogously inspired by
another differential calculus, which in particular is a 5D calculus.

The introduction of a differential calculus in $\kappa$-Minkowski
spacetime is not a trivial matter. For the 4D $\kappa$-Minkowski
spacetime one finds in the literature a few alternative versions of
4D differential calculus, and even~\cite{sitarz} the possibility of
a 5D differential calculus defined by the following commutation
relations
\begin{equation*}
[{\hat{x}}_0, \hat{\gamma}_4] = i \lambda \hat{\gamma}_0 ~~~
[{\hat{x}}_0, \hat{\gamma}_0] = i \lambda \hat{\gamma}_4 ~~~
[{\hat{x}}_0, \hat{\gamma}_j] = 0
\end{equation*}
\begin{equation}\label{eq:5Dcommutators}
[{\hat{x}}_j, \hat{\gamma}_4] = [{\hat{x}}_j, \hat{\gamma}_0] = -i
\lambda \hat{\gamma}_j ~~~ [{\hat{x}}_j, \hat{\gamma}_k] = i \lambda
\delta_{jk} (\hat{\gamma}_4 - \hat{\gamma}_0) ~.
\end{equation}
where, in light of the intuition that emerged from the analysis
reported in the previous section, we denoted the elements of the 5D
calculus using the notation $\hat{\gamma}$ which we intend to adopt
for the 5D-calculus-inspired translation-transformation parameters.

As shown in Ref.~\cite{k-Noether5D} this choice of transformation
parameters suggests a description of the translation-transformation
map $\Phi \rightarrow \Phi + \hat{d} \Phi$ with
\begin{equation}\label{eq:df5D}
\hat{d} \Phi = i\,(\,\hat{\gamma}^0 \hat{P}_0 + \hat{\gamma}^j
\hat{P}_j +\hat{\gamma}^4 \hat{P}_4\,)\, \Phi
\end{equation}\\
where the operators $\hat{P}_0, \hat{P}_j, \hat{P}_4$ are simply
related to the operators ${P}_0, {P}_j$ considered in the previous
section:
\begin{eqnarray}
&\hat{P}_0 = \frac{1}{\lambda} ( \sinh {\lambda P_0} +
\frac{\lambda^2}{2} \vec{P}^2 e^{ \lambda P_0} )\nonumber\\
&\hat{P}_i = P_i e^{ \lambda P_0}\nonumber\\
&\hat{P}_4 = \frac{1}{\lambda} ( \cosh {\lambda P_0} -1 -
\frac{\lambda^2}{2} \vec{P}^2 e^{ \lambda P_0} ) \label{eq:Phat}
\end{eqnarray}

Taking into account the coproducts of the operators $\hat{P}_0,
\hat{P}_j, \hat{P}_4$ (which one easily obtains from those of the
${P}_0, {P}_j$), and the following useful results on the commutation
relations between transformation parameters and
time-to-the-right-ordered exponentials
\begin{eqnarray}\label{eq:planewave}
&e^{i\vec{k} \cdot \vec{\hat{x}}}e^{- i k_0 {\hat{x}}_0}\,
\hat{\gamma}_0 = \left( (
 \lambda \hat{P}_0+  e^{ -\lambda P_0} )  \hat{\gamma}_0 + \lambda
 \hat{P_i} \hat{\gamma}_i
+(\lambda \hat{P}_4+1-  e^{ -\lambda P_0} )
 \hat{\gamma}_4 \right) e^{i\vec{k} \cdot \vec{\hat{x}}}e^{- i k_0 {\hat{x}}_0}\nonumber\\
&e^{i\vec{k} \cdot \vec{\hat{x}}}e^{- i k_0 {\hat{x}}_0}\,
\hat{\gamma}_i = \left(
 \lambda e^{ -\lambda P_0} \hat{P}_i \hat{\gamma}_0 + \hat{\gamma}_i - \lambda e^{ -\lambda P_0} \hat{P}_i
  \hat{\gamma}_4 \right) e^{i\vec{k} \cdot \vec{\hat{x}}}e^{- i k_0 {\hat{x}}_0}\nonumber\\
&e^{i\vec{k} \cdot \vec{\hat{x}}}e^{-i k_0 {\hat{x}}_0}\,
\hat{\gamma}_4 = \left( \lambda \hat{P}_0  \hat{\gamma}_0 + \lambda
 \hat{P}_i \hat{\gamma}_i + (\lambda \hat{P}_4+1)
 \hat{\gamma}_4 \right) e^{i\vec{k} \cdot \vec{\hat{x}}}e^{-i k_0
 {\hat{x}}_0} ~,\nonumber
\end{eqnarray}
one easily verifies~\cite{k-Noether5D} that the differential
$\hat{d} \Phi$ defined in (\ref{eq:df5D}) satisfies the Leibniz
rule:
\begin{equation}\label{liebn}
\hat{d} (\Phi \Psi) = \Phi (\hat{d}  \Psi) + (\hat{d} \Phi ) \Psi\,.
\end{equation}

The specific action we considered in Ref.~\cite{k-Noether5D} for the
Noether analysis described a free massive scalar field,
\begin{eqnarray}\label{eq:AzioneMassiva}
S[\Phi]&=&\int d^4\hat{x} \mathcal{L}[\Phi({\hat{x}})]\nonumber\\
\mathcal{L}[\Phi({\hat{x}})]&=&\frac{1}{2} \left(\Phi({\hat{x}})
\,C_\lambda\,\Phi({\hat{x}}) - m^2\Phi({\hat{x}})\Phi({\hat{x}})\right) ~,\\
\end{eqnarray}
so that the Klein-Gordon-like equation of motion takes the form
\begin{equation}\label{eq:5DMotionEquation}
C_\lambda(P_\mu)\,\Phi \equiv \left[\left(\frac{2}{\lambda} \sinh
{\frac{\lambda}{2} P_0}\right)^2-e^{\lambda
P_0}\vec{P}^2\right]\Phi=m^2\Phi ~.
\end{equation}

We can now analyze the variation of the Lagrangian density under our
5 parameter transformation. Following the same procedure of analysis
described in previous section one obtains
\begin{eqnarray}
0=\delta\mathcal{L}=\frac{1}{2}  \left( \delta\Phi\, C_\lambda \Phi
+\Phi\, C_\lambda\, \delta\Phi -
m^2\delta\Phi\,\Phi-m^2\Phi\,\delta\Phi\right)= \nonumber \\
= -\frac{1}{2} \left\{e^{\frac{\lambda
P_0}{2}}\tilde{P}^0\left[\left(\frac{2}{\lambda}+\lambda m^2-\frac{e^{\lambda P_0}}{\lambda}\right)\Phi\,\delta\Phi-\Phi \frac{e^{-\lambda P_0}}{\lambda}\delta\Phi\right]+ \right. \nonumber \\
\left. \hat{P}^i\left[\Phi e^{-\lambda
P_0}\hat{P}_i\delta\Phi-\hat{P}_i\Phi\,\delta\Phi\right]\right\}\, ,
\label{variazio}
\end{eqnarray}
where again we restricted the analysis to fields that are solutions
of the equation of motion.

In (\ref{variazio}) the transformation parameters $\hat{\gamma}_A$
appear implicitly through $\delta\Phi$. It is convenient to use the
formulas (\ref{eq:planewave}) to carry all the $\hat{\gamma}_A$ to
the left side of the monomials composing the expression of
$\delta\mathcal{L}$. This allows to rewrite Eq.~(\ref{variazio}) in
the form\\
\begin{equation}\label{protodiv}
\hat{\gamma}^A\left(\,e^{\frac{\lambda
P_0}{2}}\tilde{P}^0\,J_{0A}+\hat{P}^iJ_{iA}\right)=0\,,
\end{equation}
\noindent where
\begin{subequations}
\begin{eqnarray}
J_{00}&=&\frac{1}{2}\bigg\{\left(\frac{2}{\lambda}+\lambda
m^2-\frac{e^{\lambda P_0}}{\lambda}\right)\left[(\lambda\hat{P}_0+
e^{ -\lambda P_0} )\Phi\hat{P}_0\Phi + \lambda
 P_i\Phi\hat{P_i}\Phi
+\lambda \hat{P}_0\Phi\hat{P}_4\Phi\right]+\nonumber\\
\!\!&-&\!\!(\lambda\hat{P}_0+ e^{ -\lambda P_0} )\Phi
\frac{e^{-\lambda P_0}}{\lambda}\hat{P}_0\Phi - \lambda
 P_i\Phi\frac{e^{-\lambda P_0}}{\lambda}\hat{P_i}\Phi-
\lambda \hat{P}_0\Phi\frac{e^{-\lambda P_0}}{\lambda}\hat{P}_4\Phi\bigg\}\,, \nonumber \\
\end{eqnarray}
\begin{eqnarray}
J_{0i}&=&\frac{1}{2}\bigg\{\left(\frac{2}{\lambda}+\lambda
m^2-\frac{e^{\lambda
P_0}}{\lambda}\right)\bigg[\lambda\hat{P}_i\Phi\hat{P}_0\Phi+\Phi\hat{P}_i\Phi+\lambda\hat{P}_i\Phi\hat{P}_4\Phi\bigg]+\nonumber\\
&-&\,\lambda\hat{P}_i\frac{e^{-\lambda
P_0}}{\lambda}\Phi\hat{P}_0\Phi-\Phi\hat{P}_i\frac{e^{-\lambda
P_0}}{\lambda}\Phi-\lambda\hat{P}_i\Phi\frac{e^{-\lambda
P_0}}{\lambda}\hat{P}_4\Phi\bigg\}\,, \nonumber\\
\end{eqnarray}
\begin{eqnarray}
J_{04}&=&\frac{1}{2}\bigg\{\bigg(\frac{2}{\lambda}+\lambda
m^2-\frac{e^{\lambda P_0}}{\lambda}\bigg) \cdot \nonumber \\
&& \qquad \cdot\bigg[(\lambda\hat{P}_4+1- e^{ -\lambda P_0}
)\Phi\hat{P}_0\Phi - \lambda  P_i\Phi\hat{P_i}\Phi
+(\lambda\hat{P}_4+1)\Phi\hat{P}_4\Phi\bigg]+\nonumber\\
&& \qquad - (\lambda\hat{P}_4+1- e^{-\lambda P_0} )\Phi
\frac{e^{-\lambda
P_0}}{\lambda}\hat{P}_0\Phi +\nonumber \\
&& \qquad+\lambda
 P_i\Phi\frac{e^{-\lambda P_0}}{\lambda}\hat{P_i}\Phi-
 (\lambda\hat{P}_4+1)\Phi\frac{e^{-\lambda
P_0}}{\lambda}\hat{P}_4\Phi\bigg\}\,.
\end{eqnarray}
\end{subequations}

It is easy to verify that the charges obtained by spatial
integration of the $J_{00}$,$J_{0i}$,$J_{04}$ are time independent.
For this task it is useful to first notice that our Noether analysis
automatically brought in play the operator
\begin{equation}
\hat{\partial}_0 \equiv
e^{\frac{\lambda}{2}P_0}\tilde{P_0}=(\hat{P_0}+\hat{P_4})=\frac{e^{\lambda
P_0}-1}{\lambda} ~,
\end{equation}
which, unlike~\cite{k-Noether5D} $\hat{P_0}$, does vanish on any
time-independent field.

It is then easy to prove~\cite{k-Noether5D} that
\begin{equation}\label{eq:4Divergence}
\hat{\partial}_0 \int d^3\hat{x} \,J_{0A} = \int d^3\hat{x}
\,\hat{\partial}_0\,J_{0A}=- \int d^3\hat{x} \hat{P}^iJ_{iA} =0 ~,
\end{equation}
from which the time independence of the charges $\int d^3\hat{x}
\,J_{0A}$ follows.

Of course, it is not hard to derive an explicit time-independent
formula for the charges, which can be most conveniently
expressed~\cite{k-Noether5D} in terms of the Fourier transform
$\tilde{\Phi}(k)$ of a field $\Phi({\hat{x}})$ solution of the
equation of motion:
\begin{equation}
\left(\begin{array}{c}
\hat{Q}_0\\
\hat{Q}_i\\
\hat{Q}_4\\
\end{array}{}\right)
=-\frac{1}{2}\int d^4k \,\left|\tilde{\Phi}(k)\right|^2
\left(\begin{array}{c}
\hat{k}_0\\
\hat{k}_i\\
\hat{k}_4\\
\end{array}{}\right) \frac{(-2\tilde{k}_0e^{\frac{\lambda}{2}k_0}+\lambda
m^2)}{|-2\tilde{k}_0e^{\frac{\lambda}{2}k_0}+\lambda
m^2|}\,\delta(C_{\lambda}(k)-m^2) ~,\label{eq:Qmu}
\end{equation}
where we used again the notation $\tilde{k}_0$, introduced in the
preceding section, and we also used the notation
\begin{eqnarray*}
\{\hat{k}_0,\,\hat{k}_i,\,\hat{k}_4\}|_{k_0,\vec{k}}\equiv\big\{\frac{1}{\lambda}
( \sinh {\lambda k_0} + \frac{\lambda^2}{2} \vec{k}^2 e^{ \lambda
k_0} )\,,k_i e^{ \lambda k_0},\, \frac{1}{\lambda} ( \cosh {\lambda
k_0} -1 - \frac{\lambda^2}{2} \vec{k}^2 e^{ \lambda k_0} )\big\}~.
\end{eqnarray*}

\section{Canonical spacetimes}
\subsection{Twisted Hopf symmetry algebra and ordering issues}
The type of understanding of the symmetries of $\kappa$-Minkowski
spacetime that we reported in the previous sections was also
achieved, in our paper in Ref.~\cite{theta-Noether}, for the
symmetries of the canonical noncommutative spacetimes characterized
by the noncommutativity given in
Eq.~(\ref{thetacommutationrelation}).

The much studied~\cite{chaichaTwist,wessTwist,balaTwist}
 ``twisted" Hopf algebra of (candidate) symmetries of canonical noncommutative
spacetime can be obtained by introducing rules of ``classical
action"~\cite{aadluna} for the generators of the symmetry
algebra~\cite{theta-Noether}. In fact, observing that the fields one
considers in constructing theories in a canonical noncommutative
spacetime can be written in the form~\cite{wessDefinizio}:
\begin{equation}
\Phi(\hat{x}) = \int d^4 k \,\tilde \Phi_w(k) e^{i k \hat x}
\label{fourierW}
\end{equation}
by introducing ordinary (commutative) ``Fourier parameters" $k_\mu$,
we can associate to any given function $\Phi(\hat{x})$ a ``Fourier
transform" $\tilde{\Phi}_w(k)$, and it is customary to take this one
step further by using this as the basis for an association, codified
in a ``Weyl map" $\Omega_w$,
 between the noncommutative functions $\Phi(\hat{x})$
of interest and some auxiliary commutative functions
$\Phi^{(comm)}_w(x)$:
\begin{equation}
\Phi(\hat{x}) = \Omega_w \left(\Phi^{(comm)}_w(x)\right) \equiv
\Omega_w \left( \int d^4 k\, \tilde\Phi_w (k) e^{ikx} \right)=\int
d^4k \,\tilde\Phi_w (k) e^{ik\hat x} ~. \label{weylmap}
\end{equation}
It is easy to verify that this definition of the Weyl map $\Omega_w$
acts on a given commutative function by giving a noncommutative
function with full symmetrization (``Weyl ordering") on the
noncommutative spacetime coordinates ({\it e.g.},
$\Omega_w(e^{ikx})=e^{i k\hat x}$ and $\Omega_w (x_1
x_2^2)=\frac{1}{3}\left(\hat x_2^2 \hat x_1+\hat x_2\hat x_1 \hat
x_2 + \hat x_1 \hat x_2^2\right)$).

It is convenient\footnote{Also in the case of canonical
noncommutativity one may consider alternative ordering conventions.
The adoption of the Weyl map $\Omega_w$ essentially corresponds to
the choice of a sully symmetrized ordering convention. We have shown
in Ref.~\cite{theta-Noether}, by analyzing explicitly some
alternative choices of ordering, that our results for the charges
are independent of this choice of ordering. We shall here for
brevity not consider this issue, which readers can find discussed in
detail in  Ref.~\cite{theta-Noether}.} to use $\Omega_w$ for  our
description of the relevant  twisted Hopf algebra. This comes about
by introducing rules of  ``classical action" for the generators of
translations, space rotations and boosts:\footnote{In light of
(\ref{fourierW}) one obtains a fully general rule of action of
operators by specifying their action only on the exponentials
$e^{ik\hat x}$. Also note that we adopt a standard compact notation
for antisymmetrized indices: $A_{[\alpha \beta]}\equiv A_{\alpha
\beta}-A_{ \beta \alpha}$.}:
\begin{eqnarray}
&P_\mu^{(w)} e^{ik\hat x}&\equiv P_\mu^{(w)} \Omega_w
(e^{ikx})\equiv \Omega_w(P_\mu
e^{ikx})=\Omega_w(i\partial_\mu e^{ikx}) \label{classicalWa} \\
&M_{\mu\nu}^{(w)}e^{ik\hat x}&\equiv M_{\mu\nu}^{(w)}\Omega_w
(e^{ikx})\equiv \Omega_w(M_{\mu\nu}
e^{ikx})=\Omega_w(ix_{[\mu}\partial_{\nu]} e^{ikx}) ~.
\label{classicalWb}
\end{eqnarray}
Here the antisymmetric ``Lorentz-sector" matrix of operators
$M_{\mu\nu}$ is composed as usual by the space-rotation generators
$R_{i}^{(w)}=\frac{1}{2}\varepsilon_{ijk}M_{j k}^{(w)}$ and the
boost generators $N_{i}^{(w)}=M_{0 i}^{(w)}$. The rules of action
codified in (\ref{classicalWa})-(\ref{classicalWb}) are said to be
``classical actions according to the Weyl map $\Omega_w$" since they
indeed reproduce the corresponding classical rules of action within
the Weyl map.

It is easy to verify that the generators introduced in
(\ref{classicalWa})-(\ref{classicalWb}) satisfy the same commutation
relations of the classical Poincar\'e algebra:
\begin{eqnarray}
\left[ P_\mu , P_\nu \right]&=&0 \nonumber\\
\left[ P_\alpha,M_{\mu\nu}  \right]&=& i \eta_{\alpha [\mu} P_{\nu]} \nonumber \\
\left[M_{\mu\nu},M_{\alpha \beta}   \right]&=& i \left( \eta_{\alpha
[\nu}M_{\mu]\beta} + \eta_{\beta [\mu}M_{\nu] \alpha}  \right) ~.
\end{eqnarray}
However, they close a Hopf (rather than a Lie) algebra because the
action of Lorentz-sector generators does not comply with Leibniz
rule,
\begin{eqnarray}
M_{\mu \nu}^{(w)}\left(e^{ik\hat x}e^{i q \hat x}\right)&=&
\left(M_{\mu \nu}^{(w)} e^{ik\hat x}\right)e^{iq\hat x}+  e^{ik\hat
x} \left(M_{\mu \nu}^{(w)} e^{iq\hat
x}\right)-\frac{1}{2}\theta^{\alpha\beta}\left[ \eta_{\alpha
[\mu}\left(P_{\nu]}^{(w)}
e^{ik\hat x}\right)\cdot \right.\nonumber\\
&&\left.\left( P_\beta^{(w)} e^{iq\hat x} \right)
+\left(P_\alpha^{(w)} e^{ik\hat x}\right) \eta_{\beta [\mu}
\left(P_{\nu]}^{(w)} e^{iq\hat x}\right)\right]
~,\label{actiononexponentialsM}
\end{eqnarray}
as one easily verifies using the fact that from
(\ref{thetacommutationrelation}) it follows that
\begin{equation}
e^{ik\hat x}e^{iq\hat x}=e^{i(k+q)\hat x} e^{-\frac{i}{2}k^\mu
\theta_{\mu\nu}q^\nu}\equiv \Omega_w(e^{i(k+q) x}
e^{-\frac{i}{2}k^\mu
\theta_{\mu\nu}q^\nu}).\label{ProdottoDiEsponenziali}
\end{equation}

For the translation generators instead Leibniz rule is satisfied,
\begin{eqnarray}
P_\mu^{(w)} \left(e^{ik\hat x}e^{i q \hat
x}\right)&=&\left(P_\mu^{(w)}  e^{ik\hat x}\right)e^{iq\hat x} +
e^{ik\hat x}\left( P_\mu^{(w)} e^{iq\hat x}\right) ~,
\label{actiononexponentialsP}
\end{eqnarray}
as one could have expected from the form of the commutators
(\ref{thetacommutationrelation}) which is evidently compatible with
classical translation symmetry (while, for observer-independent
$\theta^{\mu\nu}$, it clearly requires an adaptation of the Lorentz
sector.)

In the relevant literature observations of the type reported in
(\ref{actiononexponentialsM}) and (\ref{actiononexponentialsP}) are
often described symbolically in the following way
\begin{eqnarray}
\Delta P_\mu^{(w)}&=&P_\mu^{(w)} \otimes 1
+  1 \otimes P_\mu^{(w)}   \nonumber \\
\Delta M_{\mu \nu}^{(w)}&=& M_{\mu \nu}^{(w)} \otimes 1 +  1 \otimes
M_{\mu \nu}^{(w)}-\frac{1}{2}\theta^{\alpha\beta}\left[ \eta_{\alpha
[\mu}
P_{\nu]}^{(w)}\otimes P_\beta^{(w)} +P_\alpha^{(w)}\otimes \eta_{\beta [\mu}P_{\nu]}^{(w)}\right],\nonumber\\
\label{Coproducts}
\end{eqnarray}
where $\Delta$ is the ``coproduct".

All these results can be expressed in the language of twisted Hopf
algebras: the algebra that we have just obtained is the  one
resulting~\cite{theta-Noether} from the deformation of  the
classical Poincar\'e algebra  by the twist element:
\begin{equation}
\mathcal F = e^{\frac{i}{2} \theta^{\mu\nu}P_\mu^{(w)} \otimes
P_\nu^{(w)}}. \label{twistelement}
\end{equation}

\subsection{Noncommutative transformation parameters}
In Ref.~\cite{theta-Noether} we provided a description of symmetry
transformations in canonical spacetime that follows the same
strategy already here described in the previous sections devoted to
$\kappa$-Minkowski spacetime. We wrote the symmetry-transformation
map $\Phi \rightarrow \Phi + d \Phi$ in terms of the generators
$P_\mu^{(w)},M_{\mu\nu}^{(w)}$ and of some noncommutative
transformation parameters $\gamma_\mu$,$\omega_{\mu\nu}$,
\begin{equation}
df({\hat{x}})=i \left[\gamma^\alpha_{(w)} P_\alpha^{(w)} +
\omega^{\mu\nu}_{(w)}M_{\mu\nu}^{(w)}\right]f({\hat{x}}) ~,
\label{tuttodftheta}
\end{equation}
and we assumed that the transformation parameters should still act
on the spacetime coordinates by simple (associative, but possibly
noncommutative) multiplication.

Imposing Leibniz rule on the $df({\hat{x}})$ of
Eq.~(\ref{tuttodftheta}) one finds:
\begin{eqnarray}
&&\left[
[f({\hat{x}}),\gamma^\alpha_{(w)}]+\frac{1}{2}\omega^{\mu\nu}_{(w)}(\theta_{[\mu}\,^\alpha
 \delta_{\nu]}\,^\rho + \theta^\rho\,_{[\mu}
  \delta_{\nu]}\,^\alpha)\left(P_\rho^{(w)} f({\hat{x}})\right)
     \right]P_\alpha^{(w)} g({\hat{x}})+\nonumber \\
&&+[f({\hat{x}}),\omega^{\mu\nu}_{(w)}]M_{\mu\nu}^{(w)}g({\hat{x}})=0
~,
\end{eqnarray}
which amounts (by imposing that the term proportional to
$P_\alpha^{(w)} g({\hat{x}})$ and the term proportional to
$M_{\mu\nu}^{(w)}g({\hat{x}})$ be separately null) to the following
requirements
\begin{eqnarray}
\left[f({\hat{x}}),\gamma^\alpha_{(w)} \right]&=&
-\frac{1}{2}\omega^{\mu\nu}_{(w)}(\theta_{[\mu}\,^\alpha
\delta_{\nu]}\,^\rho
+ \theta^\rho\,_{[\mu} \delta_{\nu]}\,^\alpha)P_\rho^{(w)} f({\hat{x}}) \nonumber \\
\left[f({\hat{x}}),\omega^{\mu\nu}_{(w)}\right]&=& 0 ~.
\label{CommParamF}
\end{eqnarray}
And these requirements imply the following properties of the
transformation parameters
\begin{eqnarray}
\left[{\hat{x}}^\beta,\gamma^\alpha_{(w)} \right]&
=&-\frac{i}{2}\omega^{\mu\nu}_{(w)}(\theta_{[\mu}\,^\alpha
\delta_{\nu]}\,^\beta
+ \theta^\beta\,_{[\mu} \delta_{\nu]}\,^\alpha) \label{nopureLorA} \\
\left[{\hat{x}}^\beta,\omega^{\mu\nu}_{(w)}\right]&=& 0 ~.
\label{nopureLorB}
\end{eqnarray}
As in the $\kappa$-Minkowski case, also here in considering
canonical spacetimes we are encountering a restriction on the type
of symmetry transformations that are admissible. Specifically in the
case of canonical spacetimes there cannot be any pure Lorentz-sector
transformation: according to (\ref{nopureLorB},) whenever
$\omega^{\mu\nu}_{(w)}\neq 0$ then also $\gamma^\mu_{(w)}\neq 0$.
Lorentz-sector transformations are only allowed in combination with
some component of translation transformations.

\subsection{Conserved charges}
In Ref.~\cite{theta-Noether} we verified that the description of
symmetry transformations provided in the preceding subsection was
appropriate for the Noether analysis of theory of scalar massless
fields governed by the equation of motion
\begin{equation}
\square \Phi({\hat{x}})\equiv P_\mu^{(w)} P^\mu_{(w)}
\Phi({\hat{x}}) =P_\mu^{(1)} P^\mu_{(1)}
\Phi({\hat{x}})=0.\label{equationofmotion}
\end{equation}
For the laws of transformation of the fields we of course adopted
\begin{equation}
\delta\Phi=-d\Phi=-i \left[\gamma^\alpha_{(w)} P_\alpha^{(w)} +
\omega^{\mu\nu}_{(w)}M_{\mu\nu}^{(w)}\right]\Phi({\hat{x}}) ~,
\label{VariazPhi}
\end{equation}
and we considered the following action:
\begin{equation}
S=\frac{1}{2}\int d^4{\hat{x}}
\,\Phi({\hat{x}})\square\Phi({\hat{x}}) ~, \label{action}
\end{equation}
which indeed generates the equation of motion
(\ref{equationofmotion}) and is invariant~\cite{theta-Noether} under
the transformation (\ref{VariazPhi})

 Focusing on fields that are solutions of the equation of motion,
and using the commutation relations (\ref{CommParamF}) between
transformation parameters and spacetime coordinates,
 one finds~\cite{theta-Noether} that the variation of the action
 can be written in the form
\begin{eqnarray}
\delta S &=& \frac{1}{2} \int d^4 {\hat{x}} \, \Phi({\hat{x}})
\square \delta \Phi({\hat{x}}) =  \frac{1}{2} \int d^4 {\hat{x}}
\,P_\mu^{(w)} \left[ \Phi({\hat{x}}) P^\mu_{(w)} \delta
\Phi({\hat{x}})
-  (P^\mu_{(w)}\Phi({\hat{x}})) \delta \Phi({\hat{x}}) \right] \nonumber\\
&=& - i \int d^4 {\hat{x}} \left( \gamma_\nu^{(w)} P_\mu^{(w)}
T^{\mu \nu} + \omega^{\rho \sigma}_{(w)} P_\mu^{(w)} J^\mu_{\rho
\sigma}    \right),
\end{eqnarray}
where
\begin{eqnarray}
T^{\mu \nu} &=& \frac{1}{2} \left ( \Phi({\hat{x}}) P^\mu_{(w)}
P^\nu_{(w)} \Phi({\hat{x}})
-  (P^\mu_{(w)} \Phi({\hat{x}})) P^\nu_{(w)} \Phi({\hat{x}})\right)  ,\nonumber \\
J^\mu_{\rho \sigma} &=& \frac{1}{2} \left( \Phi({\hat{x}})
P^\mu_{(w)} M_{\rho \sigma}^{(w)} \Phi({\hat{x}}) - ( P^\mu_{(w)}
\Phi({\hat{x}}) ) M_{\rho \sigma}^{(w)} \Phi({\hat{x}}) \right)
-\frac{1}{4} (\theta_{[\rho}\,^\nu \delta_{\sigma]}\,^\lambda +  \\
&& + \theta^\lambda\,_{[\rho} \delta_{\sigma]}\,^\nu) \left[
(P_\lambda^{(w)} \Phi ({\hat{x}})) P^\mu_{(w)} P_\nu^{(w)} \Phi
({\hat{x}}) - (P^\mu_{(w)} P_\lambda^{(w)} \Phi ({\hat{x}}))
P_\nu^{(w)} \Phi ({\hat{x}})  \right] ~.\nonumber
\end{eqnarray}
And we verified~\cite{theta-Noether} explicitly that the charges
obtained by spatial integration\footnote{We pose $\int d^3 {\hat{x}}
e^{ik^i {\hat{x}}_i}=\delta^{(3)}(\vec k)$.} of the
$T^{0}_{\nu}$,$J^0_{\rho \sigma} $,
\begin{equation}
Q_\mu ({\hat{x}}_0)  = \int d^3 {\hat{x}} \, T^0_\mu , \qquad
K_{\rho \sigma} ({\hat{x}}_0) = \int d^3{\hat{x}} \, J^0_{\rho
\sigma} .
\end{equation}
 are time-independent and can be conveniently written
in terms of the Fourier transform $\tilde\Phi_{(w)}(k)$ of a field
$\Phi({\hat{x}})$ solution of the equation of motion:
\begin{eqnarray}
 Q_\mu \!\!&=&\!\!   \int   \frac{d^4 q}{4|\vec q|} \,  \delta(q^2)\tilde{\Phi}_{(w)}(q)
 q_\mu  \left\lbrace  \tilde{\Phi}_{(w)}(-\vec q,|\vec q|)  \left( q^0 +|\vec q| \right)
 + \tilde{\Phi}_{(w)}(-\vec q,-|\vec q|)  \left( q^0 -|\vec q| \right)  \right\rbrace ~,\nonumber
\end{eqnarray}
\begin{eqnarray}
K_{\rho \sigma} \!\! &=& \!\! \int \frac{d^4 q}{-4i|\vec q|} \,
\delta(q^2) \tilde\Phi_{(w)}( q) q_{[\rho} \left\lbrace (q^0 \! + \!
|\vec q|)   \frac{\partial \tilde\Phi_{(w)}(-\vec q,|\vec
q|)}{\partial q^{\sigma]}} \! + \! (q^0 \! - \! |\vec q|)
\frac{\partial\tilde\Phi_{(w)}(-\vec q,-|\vec q|)}{\partial
q^{\sigma]}}  \right\rbrace. \nonumber
\end{eqnarray}

\section{Aside on covariance}
Our concept of noncommutative transformation parameters has proven
very powerful, but it will probably take quite some time to fully
appreciate its significance and implications. In this section we
want to contemplate the possibility of some rules of action of the
symmetry generators on the transformation parameters, just to show
that such rules of action can be introduced in a
logically-consistent way.

We start from the $\kappa$-Minkowski side and with the translation
generators. For these we assume\footnote{The translation generators
essentially measure the dependence of a quantity on the spacetime
coordinates, and even our new transformation parameters remain
coordinate independent.} that the action on transformation
parameters is trivial, just as in the commutative-spacetime limit:
\begin{equation}
 P_\mu(\alpha f(\hat{x}))=\alpha P_\mu f(\hat{x}),
\label{CommTransl}
\end{equation}
where $\alpha$ stands for a generic transformation parameter
($\alpha=\gamma_\mu, \sigma_j, \tau_k $).

In order to develop an analogous intuition for the action of
$\kappa$-Poincar\'e  space-rotation and boost generators we first
observe that, using (\ref{ActionOnExpAAA}),(\ref{ActionOnExpBBB}),
it is possible to describe these generators in the following way:
\begin{equation}
 R_k= \varepsilon_{klm}\hat{x}_l P_m,\qquad N_j=\hat{x}_j \bar{P_0} -\hat{x}_0 P_j,
\end{equation}
 where we introduced $\bar P_0 \equiv \left(1-e^{-2\lambda P_0}\right)/2 \lambda
 +\lambda |\vec P|^2/2$. This description of the generators $R_k$ and $N_j$
 involving exclusively translation generators and spacetime coordinates leads us to
 assume that the action of $R_k$ and $N_j$ on transformation parameters
 should indeed be derived using (\ref{CommTransl})
 and the rules of commutation (\ref{CommParam}) of the transformation parameters
 with the spacetime coordinates. It is easy to verify that from this procedure
 one obtains
\begin{eqnarray}
 R_k(\alpha f(\hat{x}))&=&\varepsilon_{klm} \hat{x}_l \alpha P_m f(\hat{x})
 =\alpha R_k f(\hat{x})+[\hat{x}_l,\alpha]\varepsilon_{klm}P_m f(\hat{x}) \nonumber \\
 N_j(\alpha f(\hat{x}))&=& \hat{x}_j \alpha \bar P_0 f(\hat{x})- \hat{x}_0 \alpha  P_j f(\hat{x})
 =\alpha N_j f(\hat{x})+[\hat{x}_j,\alpha]\bar P_0 f(\hat{x}) -[\hat{x}_0,\alpha]P_j f(\hat{x}), \nonumber \\
\end{eqnarray}
which can be conveniently reexpressed in the following way:
\begin{equation}
\left\{
\begin{array}{l}
\left[R_j , \gamma_\mu \right] = 0 \\
\left[R_j , \tau_k\right] = 0 \\
\left[R_j , \sigma_k\right] =  i \lambda  (\tau_j  P_k - \tau_n  P_n
\delta_{jk} )
\end{array}
\right. , \qquad \left\{
\begin{array}{l}
\left[N_j , \gamma_\mu \right] =  i \lambda \gamma_k  \delta^k_\mu  P_j  \\
\left[N_j , \tau_k\right] =  i \lambda \tau_k P_j   \\
\left[N_j , \sigma_k\right] =  i \lambda \varepsilon_{jkl} \tau_l
\bar{P}_0
\end{array}
\right. .\label{CommParamPoinc}
\end{equation}
And it is also easy to verify that (\ref{CommParamPoinc}) and
(\ref{CommTransl}) are fully compatible with the commutation
relations (\ref{CommParam}), and in this sense one might say that
those commutation relations are covariant.

A similar analysis is possible in the case of canonical
noncommutativity, but it is most easily formulated considering the
rules of commutation between transformation parameters and functions
of canonical-spacetime coordinates. Following the same reasoning
described above in considering the $\kappa$-Minkowski case, it is
also natural to assume a trivial action of translation generators on
the transformation parameters in canonical spacetime:
\begin{equation}
 P^{(w)}_\mu \left(\alpha f(\hat{x})\right)=\alpha P^{(w)}_\mu f(\hat{x})
\end{equation}
where, as before, $\alpha$ is a generic transformation parameter
($\alpha=\gamma_\mu,\omega_{\mu\nu}$).

For the generators $M^{(w)}_{\mu\nu}$, introduced in our description
of Lorentz-sector symmetries of canonical spacetimes, we are not
aware of any straightforward description in terms of spacetime
coordinates and translation generators. We can still propose rules
of action of the $M^{(w)}_{\mu\nu}$ generators on the transformation
parameters by posing
\begin{equation}
 M^{(w)}_{\mu\nu}\left(\alpha f(\hat{x})\right)=[M^{(w)}_{\mu\nu},\alpha]f(\hat{x})
 +\alpha M^{(w)}_{\mu\nu}f(\hat{x}) ~,
\end{equation}
and requiring ``covariance" of the commutators (\ref{CommParamF}),
{\it i.e.}
\begin{equation}
 M^{(w)}_{\mu\nu}[\alpha,f(\hat{x})]=[M^{(w)}_{\mu\nu},\alpha]f(\hat{x})+[\alpha,M^{(w)}_{\mu\nu}f(\hat{x})].
\end{equation}
Following this strategy one easily obtains
\begin{equation}
 [M_{\mu\nu}^{(w)},\omega_{\rho\sigma}]=0,\qquad [M_{\mu\nu}^{(w)},\gamma^\alpha]
 = \frac{i}{2} \omega^{\rho \sigma} \left( {\theta_{[\rho}}^\alpha \delta_{\sigma]}^\beta - {\theta_{[\rho}}^\beta \delta_{\sigma]}^\alpha \right) \eta_{\beta [\nu} P_{\mu]}~.
\end{equation}

%

\end{document}